\documentclass[12pt]{article}

\usepackage{graphicx}
\usepackage{color}
\usepackage[super]{natbib}  
\usepackage{authblk}
\usepackage{comment}
\usepackage{tablefootnote}

\usepackage[font={small},textfont={onehalfspacing}, bf]{caption}  
\usepackage[onehalfspacing]{setspace}
\DeclareCaptionLabelSeparator{bar}{ $|$ }
\def\msun{M$_{\odot}$}
\def\degs{\ifmmode ^{\circ}\else$^{\circ}$\fi}
\def\amin{\ifmmode ^{\prime}\else$^{\prime}$\fi}
\def\asec{\ifmmode ^{\prime\prime}\else$^{\prime\prime}$\fi}
\def\fss{\hbox{$.\!\!^{\rm s}$}}        
\def\farcs{\hbox{$.\!\!^{\prime\prime}$}}  
\def\h{$^{\rm h}$}
\def\m{$^{\rm m}$}

\newcommand{\gp}{\mbox{$g^\prime$}}
\newcommand{\rp}{\mbox{$r^\prime$}}
\newcommand{\ip}{\mbox{$i^\prime$}}
\newcommand{\zp}{\mbox{$z^\prime$}}

\newcommand{\ul}[1]{\underline{#1}}
\newcommand{\xmm}{{XMM-{\it Newton}}}
\newcommand{\xmmsss}{\hbox{4XMM\,J052015.1$-$654426}}
\newcommand{\HP}{\mbox{[HP99]\,159}}

\title{\bf A helium-burning white dwarf binary as a supersoft X-ray source}

\author[1]{~J. Greiner}
\author[1]{C. Maitra}
\author[1]{F. Haberl}
\author[1]{R. Willer}
\author[1]{J.M. Burgess}
\author[2, 3]{N. Langer}
\author[4]{~~J. Bodensteiner}
\author[5, 6, 7]{D.A.H. Buckley}
\author[5]{I.M. Monageng}
\author[8]{A. Udalski}
\author[9]{~~~~~~~~~~~H. Ritter}
\author[10]{K. Werner}
\author[11]{P. Maggi}
\author[12]{R. Jayaraman}
\author[12]{R. Vanderspek}
\affil[1]{Max-Planck Institut f\"ur Extraterrestrische Physik, 85748 Garching,
  Giessenbachstr. 1, Germany}
\affil[2]{Argelander-Institut f\"ur Astronomie, Universit\"at Bonn,
  Auf dem H\"ugel 71, 53121 Bonn, Germany}
\affil[3]{Max-Planck-Institut für Radioastronomie, Auf dem H\"ugel 69, 53121, Bonn, Germany}
\affil[4]{ESO -- European Organisation for Astronomical Research in the Southern Hemisphere, 85748 Garching, Karl-Schwarzschild-Str. 2, Germany}
\affil[5]{South African Astronomical Observatory, PO Box 9, Observatory Rd, Observatory 7935, South Africa}
\affil[6]{Department of Astronomy, University of Cape Town, Private Bag X3, Rondebosch 7701, South Africa}
\affil[7]{Department of Physics, University of the Free State, PO Box 339, Bloemfontein 9300, South Africa}
\affil[8]{Astronomical Observatory, University of Warsaw, Al. Ujazdowskie 4, 00–478 Warszawa, Poland}
\affil[9]{Max-Planck Institut f\"ur Astrophysik, 85748 Garching,
  Karl-Schwarzschild-Str. 1, Germany}
\affil[10]{Institut f\"ur Astronomie und Astrophysik, Kepler Center for Astro and Particle Physics, Universit\"at T\"ubingen, 72076 T\"ubingen, Sand 1, Germany}
\affil[11]{Universit\'e de Strasbourg, CNRS,
  Observatoire astronomique de Strasbourg, UMR 7550, F-67000 Strasbourg, France}
\affil[12]{Department of Physics and Kavli Institute for Astrophysics and
  Space Research, Massachusetts Institute of Technology, 
  70 Vassar St., Cambridge, MA 02139, USA}

\begin{document}

\maketitle

\newpage

\noindent {\bf
Type Ia supernovae are cosmic distance indicators\citep{Elias+1985, Riess+1996},
and the main source of iron in the Universe\citep{Hoyle1946, Burbidge+1957},
but their formation paths are still debated.
Several dozen supersoft X-ray sources, in which a white dwarf accretes
hydrogen-rich matter from a non-degenerate donor star, have been
observed\citep{Greiner2000} and suggested as Type Ia supernovae
progenitors\citep{WhelanIben1973, IbenTutukov1994, YoonLanger2003, Wang+2009}.
However, observational evidence
for hydrogen, which is expected to be stripped off the donor star during the
supernova explosion\citep{Wheeler+1975}, is lacking. Helium-accreting white
dwarfs, which would circumvent this problem, have been predicted
for more than 30 years\citep{IbenTutukov1994, Kawai+1988, IbenTutukov1989},
also including their appearance as supersoft X-ray sources,
but have so far escaped detection.
Here we report a supersoft X-ray source
with an accretion disk whose optical spectrum is completely dominated by
helium, suggesting that the donor star is hydrogen-free. We interpret the
luminous and supersoft X-rays as due to helium burning near the surface of the
accreting white dwarf. 
The properties of our system provides evidence for extended pathways towards
Chandrasekhar mass explosions based on helium accretion, in particular for
stable burning in white dwarfs at lower accretion rates than expected so far.
This may allow to recover the population of the sub-energetic so-called
Type\,Iax supernovae, up to 30\% of all Type\,Ia supernovae\citep{Foley+2013},
within this scenario.
}

\newpage

\begin{figure}[!ht]
   \includegraphics[width=0.99\textwidth]{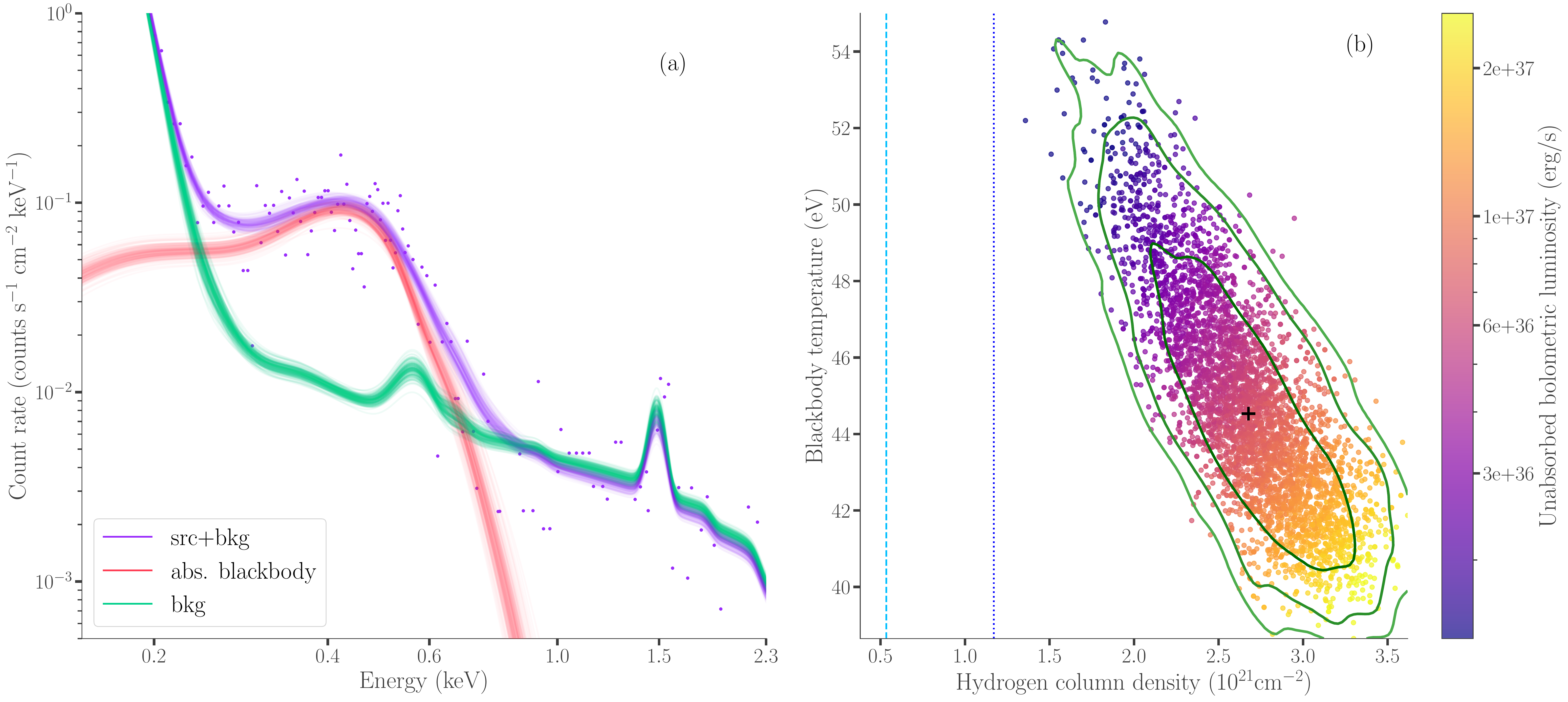}
  \caption{{\bf X-ray temperature and luminosity constraints of \HP.}
    A \texttt{3ML} fit of an absorbed blackbody model to the \xmm\
    spectrum (a), with a simultaneous fit of the background with linked parameters
    (see Methods) provides a good fit. Purple symbols and line show the
    source+background data and model, red the source only, and
    green the total background. The individual background components
    are shown in Extended Data (ED) Fig. \ref{fig:Xraybkg}. Panel (b) shows
    the posterior distribution in the  temperature vs.
    hydrogen column density plane of the spectral fit (a).
    Each dot represents one model realization.
    The colour coding represents the unabsorbed bolometric luminosity
    assuming a distance of 50\,kpc.
    1$\sigma$, 2$\sigma$ and 3$\sigma$ confidence contours are overplotted
    in green.
    Vertical dashed lines mark the Galactic foreground absorption
    and the sum of Galactic and total LMC absorption. 
  }
\label{fig:Xrayspec}
\end{figure}

\begin{figure}[!ht]
  \centering
  \vspace{-0.3cm}
  \includegraphics[viewport=10 0 530 353, clip, width=11.8cm]{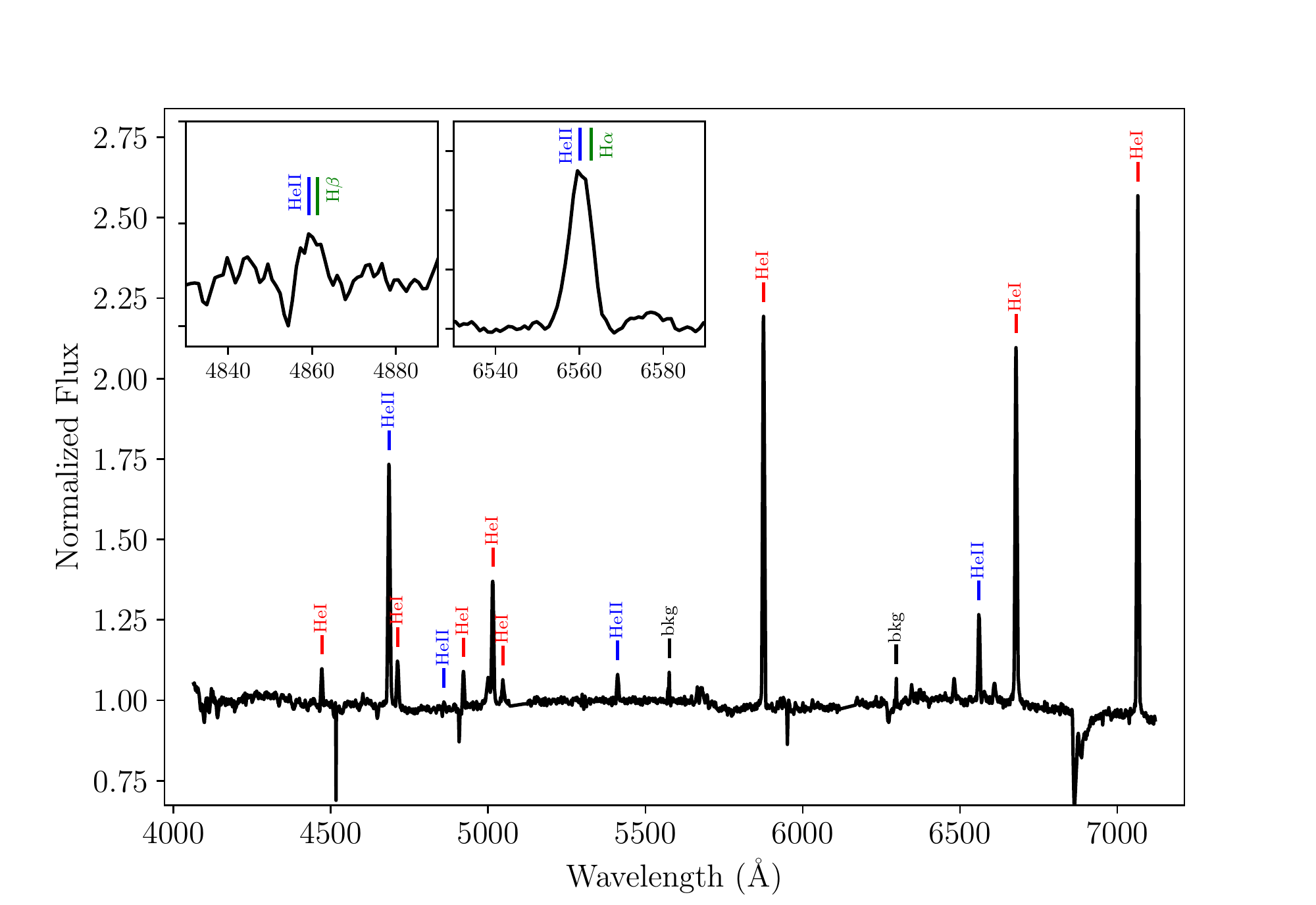}
  \vspace{-0.3cm}
  \caption[Lowres]{{\bf Low-resolution optical spectrum of \HP.}
  Flux-normalized optical spectrum taken with the SALT/RSS spectrograph
  at 2020-08-14 03:44 UT (mid-time of 1200 s exposure), with the major emission
  lines labeled ('bkg' labels residuals of removing sky lines).
  The three absorption features apart from
  the B-band are also due to residuals of sky line removal.
  The inserts demonstrate that the 4860 \AA\ and 6560 \AA\ lines are
  due to He II, and not hydrogen.
  \label{fig:lowres}}
\end{figure}

The X-ray source \HP\citep{HaberlPietsch1999} has been seen since the early
1990s with ROSAT, XMM-{\it Newton} (\xmmsss) and recently
eROSITA (eRASSU  J052015.3-654429) with a very soft spectrum
(effective blackbody temperature of kT = 45$\pm$3 eV, or 522$\pm$35 kK;
Fig. \ref{fig:Xrayspec}).
Using the 1\asec-accurate XMM X-ray position, we identify \HP\ with a
16 mag object at RA (2000.0) = 05\h20\m15\farcs50,
Decl. (2000.0) = $-$65\degs44\amin27\fss1.
An optical spectrum taken with the Robert Stobie Spectrograph (RSS) at
the Southern African Large Telescope (SALT) shows a wealth of emission lines
(Fig. \ref{fig:lowres}), all shifted by the Large Magellanic Cloud (LMC)
systemic velocity\citep{vanderMarel+2002} of 262.2$\pm$3.4 km s$^{-1}$,
indicating that the source is indeed situated at LMC
distance (50 kpc\citep{Pietrzynski+2019}).
Thus, the X-ray fit leads to a high bolometric luminosity of
6.8$^{+7.0}_{-3.5}$ $\times$ 10$^{36}$ erg/s.
The corresponding blackbody radius is 3700$^{+3900}_{-1900}$ km,
consistent with a white dwarf.
This classifies \HP\ as bona-fide supersoft X-ray source
\citep{Greiner+1991, Heuvel+1992, Greiner2000}.

The optical spectrum is unique, in that it shows predominantly He I and He II
emission lines (Fig. \ref{fig:lowres}). There are no indications for
Balmer lines (see inserts in Fig. \ref{fig:lowres}), no absorption lines
typical for a main-sequence star, and no indications either for C or O
as seen in Wolf-Rayet stars. The only other emission lines we identify
(ED Fig. \ref{fig:NIISII}) are 7 lines of
NII (5001.5, 5666.6, 5679.6, 6482.0, 6610.6 \AA) and SiII (6347.1, 6371.4 \AA).
While such lines are typically seen in AM CVn stars, several facts
argue against such an interpretation.
We  find no evidence
in the extracted 2D long-slit spectrum of any extended nebulous emission.
The strong continuum emission argues against an HII-like region 
of a (He-rich) planetary nebula.

\begin{figure}[!ht]
  \centering
  \includegraphics[width=12.cm]{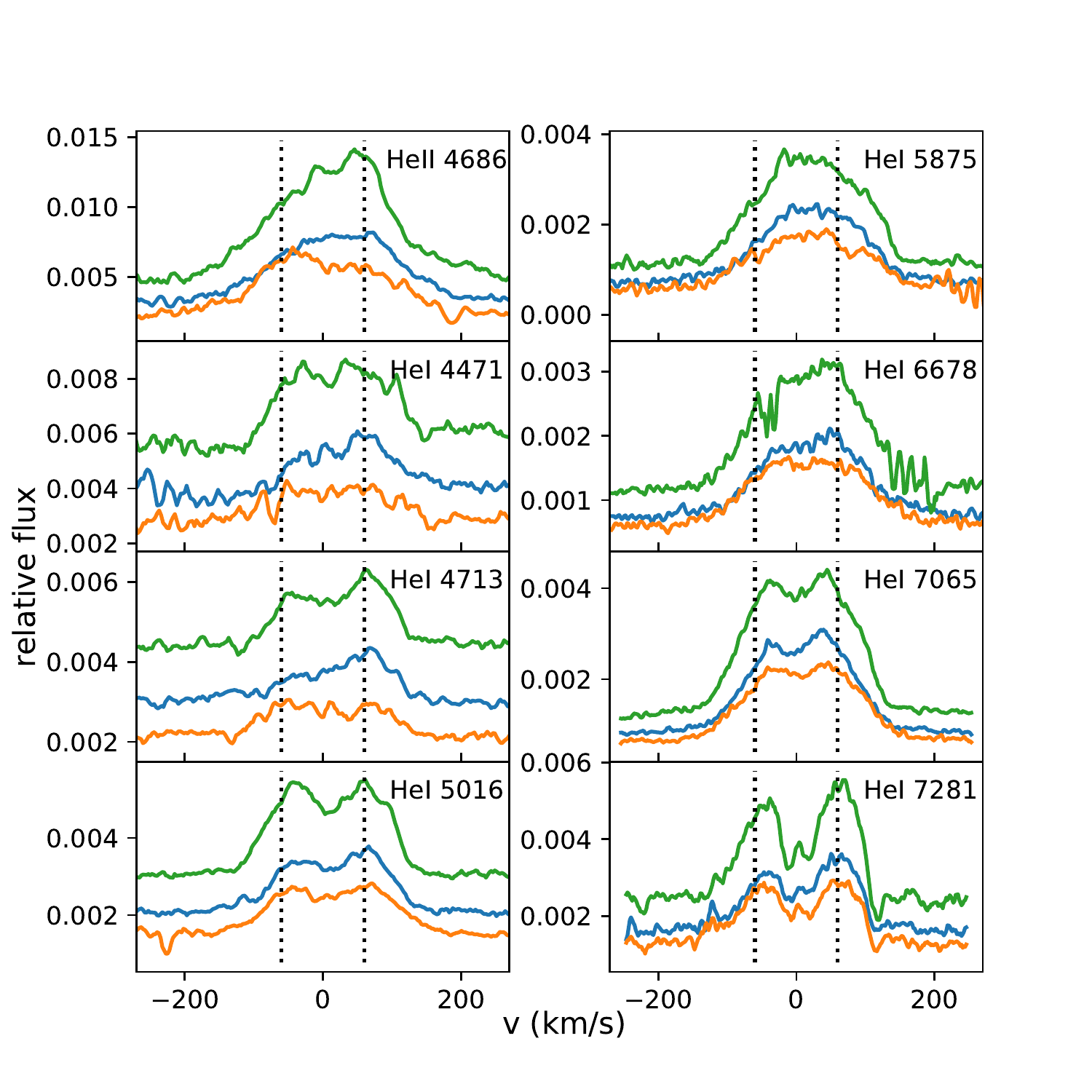}
  \vspace{-0.5cm}
\caption[Lineshape]{{\bf Double-peak shape of optical emission lines.}
  Flux-normalized optical spectra of different He lines (as labelled
  on the top right of each panel) taken with the SALT/HRS spectrograph
  at three different epochs: 15 Sep 2020 (dashed), 5 Oct 2020 (dotted)
  and 6 Oct 2020 (solid).
  The peak separation of all major lines 
  is similar, incl. that of He II. The vertical dotted
  lines indicate a peak separation of $\pm 60\, {\rm km\,s^{-1}}$.
  The relative variation of blue/red peaks is usually explained as the
  orbiting hot spot created by the impact of the accretion stream
  on the outer edge of the accretion disk.
  \label{lineshape}}
\end{figure}

High-resolution optical spectra taken at 3 epochs with the HRS
(High Resolution Spectrograph) at SALT reveal a double-peaked profile
of all lines (Fig. \ref{lineshape}), thus demonstrating their origin in
an accretion disk.
With the theoretical maximum intensity of an accretion disk line profile
coming from the area of  $\approx$0.95 of its maximum Doppler velocity,
and assuming Keplerian rotation, we infer a
projected velocity of the outer disk of
$v_K \times {\rm sin} (i) \approx 60\, {\rm km\, s^{-1}}$.
This suggests that the disk is seen at a low inclination angle,
close to face-on.
Interestingly, the He II lines have a similar profile.
The FWZI (full width at zero intensity) in the He I lines suggests a maximum
projected velocity of 120$\pm$10 km/s, with that of the He II 4686 line clearly
being different, about 200$\pm$20 km/s.

The accretion disk is not only the origin of the emission lines, but 
also of the  UV-optical-NIR continuum emission, as indicated by its
luminosity and spectral slope; 
the accreting white dwarf and the
donor are both hidden under this disk flux.
Optical photometry shows periodic variations by a factor of 1.3, with
little color variation (see ED Fig. \ref{fig:machooglelc}).
A Lomb-Scargle periodogram shows the largest power
at a period of 1.1635 days,
and a secondary lower-power peak at 2.327 days.
The folded light curve for this longer period 
has a lower variance and a clear odd-even asymmetry. Phase-resolved
spectroscopy is certainly needed to firmly establish which one is
the true orbital period.

The helium-dominated accretion disk 
has two consequences: First, the donor star must be 
in an evolutionary phase where all the hydrogen is lost.
An intriguing option is a helium star donor, with
the nitrogen lines providing evidence for CNO-processed matter
from the donor.
Secondly, we
interpret the high X-ray luminosity as due to steady He burning  
in a shell on the white dwarf (accretor) surface.
Similar to the steady H-shell burning in the canonical supersoft X-ray
sources \citep{Nomoto1982, Fujimoto1982},  models of accreting white dwarfs predict a narrow range
of accretion rates, with a canonical value of $\sim$10$^{-6}$ \msun/yr, at which He-shell burning
is steady \citep{Kawai+1988, IbenTutukov1989, IbenTutukov1994, YoonLanger2003,
  Yoon+2004, Piersanti+2014, Wong+2021}.
If the accretion rate is higher, the accreted material puffs up and
forms an envelope around the WD which becomes similar to a red giant,
likely leading to common envelope evolution. If the accretion rate is
lower, burning in the accreted He-layer is unstable, i.e., first starting to oscillate
and then leading to He shell flashes that 
increase the luminosity temporarily by factors of 10 or more, on timescales
which depend on various parameters \citep{YoonLanger2004, Brooks+2016}.
Even lower accretion rates result in explosive helium burning.

While the measured X-ray temperature is exactly in the range expected
for steady He shell burning, our measured luminosity is about ten times
smaller than expected for accretion at the canonical rate.
At the same time, the historical X-ray light curve, from
{\it Einstein} (1979) and EXOSAT limits (1984--1986) to
the ROSAT detection in 1992, and the XMM-{\it Newton} and eROSITA
detections since 2019, suggests that the luminosity of \HP\ is stable
to within a factor of 5 (relative to the \xmm\ value) for nearly 50 years
(ED Fig. \ref{xraylc}).
This indicates the possibility that helium accretion at rates well below the
canonical
one (i.e., $\approx 10^{-6}$ \msun/yr) can still lead to stable helium burning.

Stable burning at low accretion rates has been suggested for the case that the accreting white dwarf 
is rapidly rotating \citep{YoonLanger2004, WongSchwab2019}. In corresponding models, 
stable burning is found\cite{Yoon+2004} down to 5$\times$10$^{-7}$ \msun/yr, 
and even for 3$\times$10$^{-7}$ \msun/yr 
when allowing for fluctuations of the burning rate of a factor three.
In the latter situation,
the X-ray luminosity at any given time may be up to a factor of three smaller, or larger, 
than the value deduced from a given accretion rate assuming strictly stationary burning.
If \HP\ were currently near a luminosity minimum, which is more likely than it being near a maximum,
its helium accretion rate could indeed be as high as  3$\times$10$^{-7}$ \msun/yr.  
While we can not exclude that the burning rate of \HP\ is oscillating with a growing amplitude, leading to
instability, the expected short timescale of the evolution renders this unlikely. 

A lower than the canonical burning rate is consistent with our optical spectra.
If the accretion rate
in \HP\ would be significantly higher, a wind from the white dwarf
is expected\citep{KatoHachisu1999}. This would manifest itself
with emission lines, broadened by the wind velocity (of order thousands
km/s).
Such broadened lines are not detected.

We have the following constraints on the mass of the He star:
For initial He-star masses above $\approx$1 \msun, long-term stable evolution
has been found\citep{Wang+2009}.
The maximum possible initial mass depends on the assumptions concerning
the wind. The present mass could be smaller than that.
A rough upper bound on the present mass could be derived
using the constraint that its luminosity is obviously smaller
than that of the accretion disk.
A helium star luminosity below $\approx 1000\,$L$_{\odot}$
implies\citep{Langer+1989} that the 
current mass of the helium star is smaller than $\approx 2\,$M$_{\odot}$.

An orbital period of (1.16\,d) 2.32\,d suggests that the He star fills
its Roche lobe radius of $\approx$(3) 4 $R_{\odot}$, being about a factor of
10 larger than on the He main sequence.
In this picture, as long as the mass of the He star donor is larger than that
of the white dwarf accretor, mass transfer will proceed on the thermal
timescale ($\approx$10$^5$...10$^6$ yrs), reducing the separation of the stars.
Indeed, for He stars in the  0.8--2 \msun\ range (corresponding to
initial masses on the main-sequence of 4--8 \msun), this thermal timescale
mass transfer\citep{DelgadoThomas1981}
(during their sub-giant or giant phases) is predicted to reach
rates of order  10$^{-7}$...10$^{-5}$ \msun/yr, allowing
for stable He burning.
After mass ratio inversion, the mass transfer rate drops and the
binary widens. This may lead to the weak He-shell
flash regime, consistent with [HP99] 159.

Various scenarios of white dwarfs accreting matter from a helium star
companion have been suggested to lead to Type\,Ia supernovae. At the
lowest accretion
rates, helium can pile up on the white dwarf and lead to a sub-Chandrasekhar
mass explosion after a critical amount of mass has been accumulated. However,
in \HP\, the X-ray emission implies continuous burning of the accreted
matter, and consequently a continuous growth of the white dwarf mass.
For this case,
it has been suggested that the white dwarf undergoes a Type\,Ia
supernova explosion once the Chandrasekhar mass is reached. A standard
Type\,Ia explosion may strip 2$\dots$5\% of the mass of the helium
star\citep{Liu+2013}, of which no signature has been observed so far.
However, it has been suggested that Chandrasekhar mass WDs may undergo
sub-energetic deflagrations\citep{Kromer+2013},
leading to subluminous so-called Type\,Iax supernovae, which are expected to
strip off about ten times less mass from their helium donors\citep{Zeng+2020}.
Weak helium lines have been observed in the spectra of two Type\,Iax
supernovae\citep{Foley+2013}, and a
helium donor star has been proposed for the Type\,Iax SN\,2012Z based on deep
pre-explosion imaging\citep{McCully+2014}.
The recent detection of helium in the bright
Type\,Ia SN\,2020eyj \citep{Kool+2022} indicates that helium donors may
also sometimes trigger energetic white dwarf explosions.

While we do not know whether \HP\ will evolve into a Type\,Ia supernova, its
properties provide evidence for the pathway towards Chandrasekhar mass
explosions based on helium accretion being wider than thought before. Its
X-ray luminosity of $\sim 1800\,$L$_{\odot}$ corresponds to a
stationary helium accretion rate of $1.5\,10^{-7}\,$\msun/yr, for which many
models currently predict unstable burning \cite{WongSchwab2019}.
However, \HP\ appears to be
relatively stable within the last 50\,yr. Stable burning for lower accretion
rates, as perhaps enabled by rapid rotation\citep{Yoon+2004},
may allow lower mass donors to push their companion WDs to the
Chandrasekhar mass. This may allow us to recover the SN\,Iax population
within this scenario, which makes up about 30\% of all Type\,Ia
supernovae\citep{Foley+2013}.
Folding our constraint on the radius of the WD in \HP\ with a WD mass-radius
relation\citep{Rotondo2011}, we find a current WD mass of
$1.20^{+0.18}_{-0.40}$\,M$_{\odot}$, implying that \HP\ could
undergo a Type\,Iax supernova explosion in the future.

When we assume that $\approx$10\% of all Type\,Ia supernovae in our Galaxy
($\approx$10$^{-3}$ per year\citep{Wang+2009})
follow the path of helium accretion leading to Type\,Iax explosions,
and adopting a lifetime of 3$\times$10$^5$ yrs (assuming 0.3 \msun\
need to be transferred at 10$^{-6}$ \msun/yr), we predict about 30 
helium accreting supersoft X-ray sources presently in the
Milky Way. Scaling with the star formation rate would yield
a handful of systems in the LMC. The detection and study of
more of these sources will likely allow us to tighten the constraints
on the single degenerate progenitor channel for Type\,Ia supernovae. 

{}

\newpage

\newpage

\newpage

\newpage

\begin{figure}[!ht]
    \includegraphics[width=16.0cm]{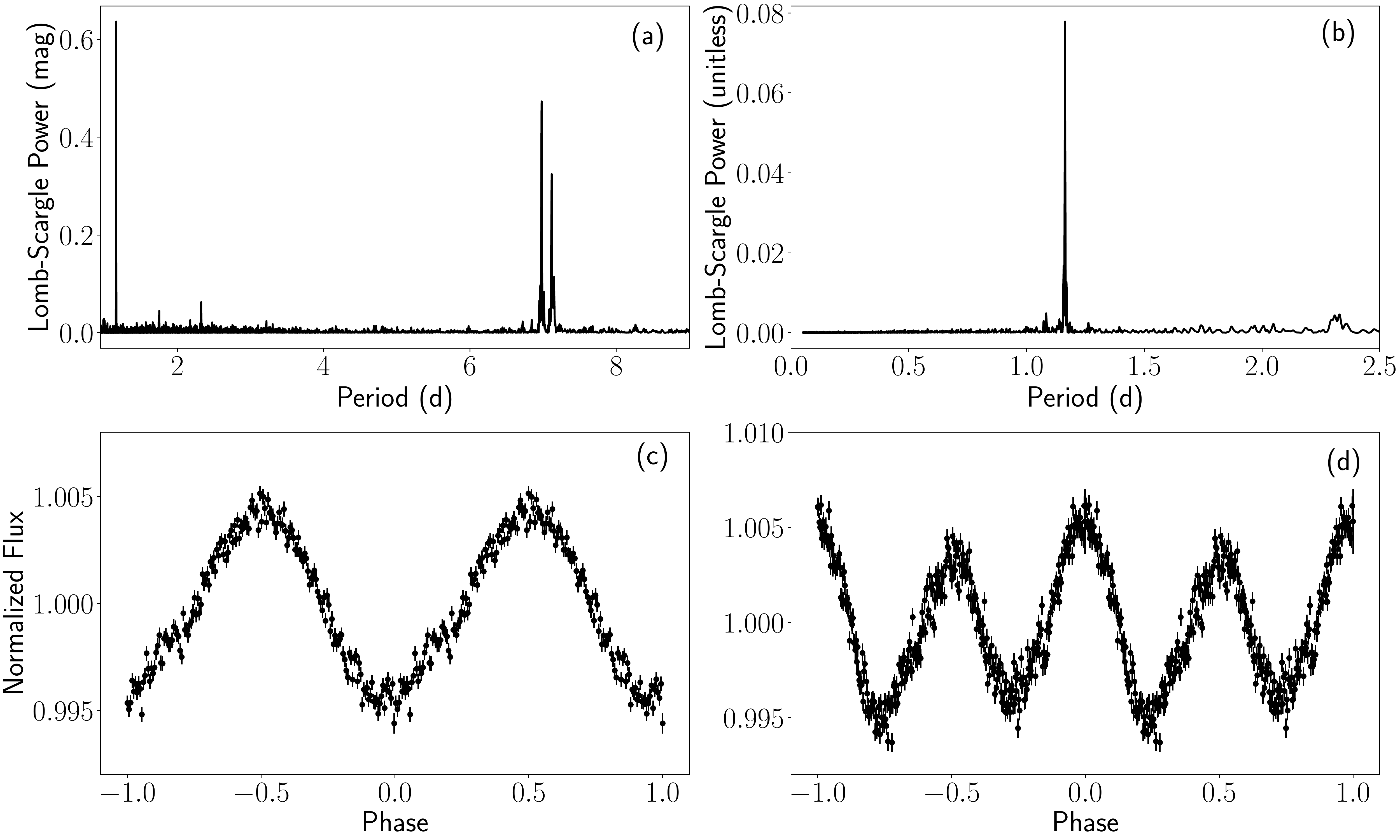}
  \vspace{-0.2cm}
  \caption[Folded lc]{{\bf Optical light variation.}
    Lomb-Scargle periodograms derived from the OGLE I-band data (a) and the TESS data (b).
    For the interpretation of the OGLE peaks see ED Tab. \ref{tab:OGLE_f}.
    The TESS data are folded with a period of P = 1.1635 days (c)
    and P = 2.327 days (d), corresponding to ephemeres of
    2459036.2885297858(14) + 1.1635*N and 2459036.288557(8) + 2.327*N
    (Barycentric Julian Date), respectively. The error bars for a
    given point represent the rms error of the individual 10 min data points
    that went into that one bin.
    While the Lomb-Scargle periodograms show the highest power at
    P = 1.1635 days, the folded light curve for the longer period (d)
    has a lower variance and a clear odd-even asymmetry.
    Lacking phase-resolved spectroscopy for a definite proof, we 
    tentatively identify P = 2.327 days as the orbital period of \HP.
  \label{fig:periodogram}}
\end{figure}

\newpage

\section*{Methods}


\noindent{\bf Optical photometry:} 

\noindent\ul{SkyMapper:}
The optical brightness, measured by SkyMapper\cite{Onken+2019}
(not simultaneously) is
\gp = 15.82$\pm$0.02 mag,
\rp = 16.04$\pm$0.02 mag,
\ip = 16.41$\pm$0.01 mag,
\zp = 16.59$\pm$0.04 mag,
and after correcting for the Galactic and LMC reddening of
E(B-V) = 0.105 mag (see below) results in an absolute
V-band magnitude of $M_V = -2.8$ mag 
(assuming a LMC distance \citep{Pietrzynski+2019} of 50 kpc).
This is about 5 mag (or a factor 2.5$^{5}=100$)  brighter than typical disks
in high-accretion rate nova-like cataclysmic variables \citep{Smak1989},
and still
15--40x brighter for a face-on disk.

\noindent\ul{OGLE:}
The region of our X-ray source was monitored regularly in the $V$ and $I$ bands
with the  Optical Gravitational Lensing Experiment
\citep[(OGLE)][]{Udalski+2008, Udalski+2015} at a cadence of 1--3 days.
Photometric
calibration is done via zeropoint measurements in photometric nights,
and color-terms have been used for both filters when transforming
to the standard $VI$ system.
The long-term lightcurve
over the 2010--2020 period shows variations
by a factor 1.3 and little color variation (see ED Fig. \ref{fig:machooglelc}).
A Lomb-Scargle periodogram identifies a period of P = 1.1635 days
with the largest power (panels (a) and (b) of Fig. \ref{fig:periodogram}),
in agreement with P = 1.163471 days listed in the
 EROS-2 catalog of LMC periodic variables
 \citep[(EROS-ID lm0454n2690)][]{Kim+2014}.
Two other strong
peaks at longer periods are aliases (see ED Tab. \ref{tab:OGLE_f}).
A much smaller peak is seen at 2.327 days (see below).

\noindent\ul{MACHO:}
The source was also covered by the MACHO project
\citep{Alcock+1999} that monitored the brightnesses of
60 million stars in the Large and Small Magellanic Clouds,
and the Galactic bulge between 1992--1999.
A visual (4500--6300 \AA) and a red filter (6300--7600 \AA) were used,
the magnitudes of which were transformed
to the standard Kron-Cousins $V$ and $R$ system, respectively,
using previously determined color-terms \citep{Greiner+2002}.

\noindent\ul{TESS:}
The Transiting Exoplanet Survey Satellite\cite{Ricker+2014} (TESS)
is an all-sky transit survey
to detect Earth-sized planets orbiting nearby M dwarfs.
It continuously observes a given region of the
sky for at least 27 days.
For sources
down to white light magnitudes $\approx 16$ mag, TESS achieves
$\approx$1\% photometric precision in single 10 min. exposures.
However, its large plate scale
(21\asec px$^{-1}$) means that care must be taken wrt. to blended sources.

\HP\
was observed during all of TESS Sectors 27--39 (except Sector 33), i.e.
from 2020 July to 2021 June.
The analysis of \HP\ is complicated by a 13\,mag star at 12\asec\
distance. Yet, the 1.16 d period found in OGLE data (which resolves
these two stars) is clearly visible in a Lomb-Scargle
periodogram of the TESS data
(Fig. \ref{fig:periodogram}) as the strongest peak by far.
There is a signal at 2.3268 d, exactly twice of the OGLE period,
at a significance of 3$\sigma$. While this is marginal, the folded
(and re-binned) light curve reveals a clear odd-even effect with smaller
variance that
leads us to believe that this is the true period, and the
strong peak at 1.16\,d is likely the first harmonic of this period. 
The small amplitude difference, at the 0.2\%-level, would explain
that this is only marginally seen in the TESS periodogram.
This period is also seen in the OGLE periodogram,
demonstrating that it is a real feature.
The phenomenon of asymmetrical maxima and minima,
known in some detached binaries\citep{Young+1972},
is unique in interacting binaries, and is especially puzzling given our
inferred near face-on geometry.

With the TESS light curve\citep{Feinstein+2019}, we also did an independent,
more sensitive
search at even shorter periods that are inaccessible to OGLE.
The TESS light curve was pre-whitened of the 1.16 d
period and 25 of its harmonics, and the Fourier transform of the
``cleaned'' data was calculated.
There are no indications for a shorter period down to $\approx$3 hrs
(Fig. \ref{fig:periodogram}).
There is also no signal at 0.538 days. This would be the
fundamental period if the 1.16 d period still were an alias with
the 1-3 days observing cadence of OGLE.
On the other hand, two additional periodicities are found,
at $P_1 = 2.635$ h and $P_2 = 1.32$ h, with significances at the
4$\sigma$ level\footnote{We assume that the noise is Gaussian and calculate
the standard deviation
in a 1500-bin window ($\pm$ 0.1 cycles/day in frequency) around any
identified peaks}. Given the
non-Poissonian nature of the light curve after pre-whitening, we do not
consider these two periods,
which are not related harmonically, to be significant enough for further
investigation.

\noindent\ul{Swift/UVOT:}
A 1061 second {\it Swift} observation was obtained on Aug. 9, 2022, starting at
23:15 UT. While not detected in X-rays (as expected, Fig. \ref{xraylc}),
we detect \HP\
in all filters of the ultraviolet-optical telescope (UVOT),
at AB magnitudes as follows:
UVW2 = 15.29$\pm$0.04 mag,
UVW1 = 15.33$\pm$0.04 mag,
U = 15.44$\pm$0.04 mag,
B = 15.73$\pm$0.04 mag,
V = 15.93$\pm$0.05 mag,
where the error is the quadratic sum of statistical and systematic error.
When added to the (non-simultaneous) measurements on the longer-wavelengths
bands (ED Fig. \ref{fig:SED}), the spectral energy distribution is
still well described by a straight powerlaw, extending from
0.2--8 $\mu$m, without any sign of the He donor.

\noindent\ul{SED modelling and extinction correction:}
The recent reddening map\citep{Skowron+21} of the LMC 
returns a much smaller reddening than previous estimates.
In addition, it provides a combined reddening value
for the Galactic foreground and the median LMC-intrinsic value,
together with a spread due to variation within the LMC.
Instead of trying a somewhat arbitrary 
extinction correction, we instead forward-fold a powerlaw
model to all the photometry from Swift/UVOT, SkyMapper,
2MASS and Spitzer.
We fit for the powerlaw slope
extinguished by a combination of Milky Way and LMC dust.
The powerlaw model fit is very good, and does not require a
more complicated spectral model (ED Fig. \ref{fig:SED}).
The best-fit values are a powerlaw slope of $\nu^{1.48\pm0.02}$,
and E(B-V) values of 0.01$\pm$0.01 for Milky Way
and 0.14$\pm$0.01 for LMC dust.
The latter is somewhat larger than the E(B-V)=0.11 mag  provided by the LMC
reddening map\citep{Skowron+21} (composed of
E(I-V)=0.08 mag to the center of the LMC
and an additional E(I-V)=0.06 mag towards the far end of the LMC).
More importantly, the slope of the spectral energy distribution
is different
from that expected for a standard accretion disk $F_\nu \propto \nu^{1/3}$
(ED Fig. \ref{fig:SED}).
This is very similar to the SEDs of other supersoft X-ray
sources like CAL 83\cite{Crampton+1987}.
The flatter slope has been interpreted as due to reprocessing of the
high-luminosity soft X-rays,
making the emission $\approx$100--1000 times larger than the
accretion luminosity \citep{PophamDiStefano1996}.

\noindent{\bf Optical spectroscopy:}

Optical  spectroscopy  of  our source  was  undertaken on the
Southern African Large Telescope (SALT).
On 14 August 2020 a 1200 s long-slit exposure was obtained using the
Robert Stobie Spectrograph \citep[(RSS)][]{Burgh+2003}
in the 4070--7100 \AA\ range (Fig. \ref{fig:lowres}). 
Three further exposures (2020 Sep. 16, Oct. 06 and 07),
using the High Resolution Spectrograph \citep[(HRS)][]{Crause+2014},
covered the 3700--5500 \AA\ and 5500--8900 \AA\ wavelength ranges.
The primary reduction, which includes overscan correction, bias subtraction
and gain correction, were carried  out  with  the  SALT  science  pipeline
\citep{Crawford2015}.

\noindent{\bf X-ray analysis:}

\noindent\ul{XMM-Newton:}
4XMM J052015.1-654426 
was covered serendipitously  in a 29~ks \xmm\
observation (ObsId 0841320101, PI: Pierre Maggi) on
2019 September 16/17. The EPIC instruments were operating in full-frame mode,
with thin and medium filters for the pn and MOS detectors, respectively.
We used the \xmm\ data analysis software \texttt{SAS} version 20.0.0 to
process these data.
Good time intervals were identified following the method described at 
\texttt{https://www.cosmos.esa.int/web/xmm-newton/sas-thread-epic-filterbackground}.
A whole field-of-view lightcurve for single-pixel events with $10000 < PI < 12000$ is
created and visually inspected for periods of flaring. A quiescent rate of
less than 0.46 cts/s is determined and a GTI file satisfying this condition is
created and used to filter the observation. After this 
filtering and given the off-axis position (8.7 arcmin) of \HP, its
resulting vignetted exposure was $\approx$11.5 ks. The
events used for the spectral analysis were filtered with the following
expression using the SAS task \texttt{evselect}:
{\tt '(PATTERN == 0) \&\& (PI in [150 : 15000]) \&\& (FLAG == 0)'}.
The SAS task \texttt{especget} was used to
extract (source and background) events from a circular region with
radius 60\asec\ centered on the position 
RA (2000.0) = 5\h20\m15\fss4, Decl. (2000.0) = $-$65\degs44\amin32\asec,
as well as to calculate RMF and ARF for these events.
The same was done with a circular region with radius 110\asec\
centered on the position
RA = 5\h20\m15\fss5, Decl. = $-$65\degs41\amin11\asec,
to be used as the background only,
after excising two point sources in that region.
In order to estimate the spectral parameters of the source,
a Bayesian approach was
implemented via \texttt{3ML}\citep{Vianello+2017, Burgess+2021}.
The analysis was restricted to the 0.2$-$2.3 keV energy band.
The background and source contribution to the detected photons were
modelled and folded through the appropriate responses to calculate posterior
distributions of the spectral parameters. The source was modelled as an
absorbed blackbody, using the \texttt{3ML} models TbAbs*Blackbody
(no separate abundances are used for the foreground Galactic and the
LMC-intrinsic absorption).
The background was
modelled as a combination of instrumental background (read noise and
fluorescence lines) and astrophysical background (Fig. \ref{fig:Xraybkg}),
as follows:
(i) a Gaussian line with normalisation, line energy and width left free to
account for the low-energy noise introduced by the read-out electronics,
(ii) a Gaussian line with line energy and width fixed representing the Al-K
fluorescent line near 1.5 keV, which is excited by particles in the camera
body,
(iii) an unabsorbed APEC model with temperature left free to vary around
0.11 keV accounting for the hot gas of the local bubble,
(iv) an APEC model with temperature allowed to vary around 0.22 keV absorbed
by the average Galactic hydrogen column in the direction of the source,
describing the contribution from the Galactic halo, and
(v) a powerlaw with fixed slope of $-1.41$, absorbed by the combined hydrogen
column of the Galaxy and the LMC in the direction of the source, arising from
unresolved AGN. The contribution of the particle background is negligible
in our spectral range.
The photons in the source extraction region were modelled by adding the source
spectrum and the background spectrum, scaled by the ratio of the extraction
areas. During the fit of the data, the parameters describing the
background models were linked.
We obtain the following best-fit values (errors at the 1$\sigma$ level):
$kT = 45\pm3$ eV, $N_{\rm H} = (2.7\pm0.4) \times 10^{21}$ cm$^{-2}$,
and an unabsorbed bolometric luminosity of
$6.8^{+7.0}_{-3.5} \times 10^{36}$ erg/s, see Fig. \ref{fig:Xrayspec}.
This implies an emission radius of 3700$^{+3900}_{-1900}$ km,
consistent with a white dwarf radius.

Apart from the possibility of the flux-oscillations due to the accretion rate
being slightly below the burning rate, two other factors may contribute
to the discrepancy of the measured vs. expected X-ray luminosity.
First, due to the accretion of pure helium, the burning proceeds via the
triple-$\alpha$ process \citep{Hansen+2004}, with log$T (K) \approx  8.4$
and $\rho \approx 1000$ g cm$^{-3}$ at the burning depth, leading to
elevated levels of carbon and oxygen. Convective
envelope mixing and subsequent wind ejection of CO-rich matter
could lead to noticeable local X-ray absorption in the emission volume.
Secondly, Non-LTE model atmospheres (as frequently used for the
supersoft phase in post-nova) usually give a higher peak
intensity\citep{RauchWerner2010} than blackbody models
(at the same temperature).
Both effects, if taken into account in future work with improved data,
would likely result in a higher X-ray luminosity
(and WD radius) than that estimated above.

\noindent\ul{eROSITA:}
\HP\ = eRASSU J052015.3-654429 was detected by
eROSITA\citep{Predehl+2021} in each of the survey scans.
Until the end of 2021, {\it eROSITA} scanned the source during
five epochs as summarised in Table.~\ref{tabobsero}.
The X-ray position was determined from the combined four eRASS surveys to be
RA (2000.0) = 05\h20\m15\fss52 and Decl. (2000.0) = $-$65\degs44\amin28\farcs9
with a $1\sigma$ statistical uncertainty of 0\farcs6.
The positional error is usually dominated by systematic
uncertainties\citep{Brunner+2022} which presently amount to 5\asec\
in pointed and 1\asec\ in scanning observations.

Due to the unprecedented energy resolution (about 56 eV at 0.28 keV),
eROSITA data are particularly sensitive to temperature changes of the
source. Thus, we decided to perform spectral fitting despite the
low number of counts. The spectral analysis was done using the
five detectors with the on-chip aluminium
filter (telescope modules 1, 2, 3, 4, and 6), avoiding the light leak
in the other two detectors \citep{Predehl+2021}.
The eSASS\citep{Brunner+2022} users version 211214 was used to process the data.
Only single-pixel events without any rejection or
information flag set were selected, using the eSASS task \texttt{evtool}. 
With the eSASS task \texttt{srctool}, a circular source region with a radius 100\asec,
centered on the coordinates RA(2000) = 5:20:16.6, Decl.(2000) =$-$65:44:27
was defined to select source events.
A background region of the same size and shape centered on
RA(2000) = 05\h21\m09\fss4, Decl.(2000) = $-$65\degs46\amin00\asec\ was defined,
so as to lie
at the same ecliptic longitude as the source region, and hence in the
scanning direction
of eROSITA. The corresponding ARF and RMF files were created by the same eSASS
task. Spectra were constructed by combining all events within the
respective regions for each of the 5 epochs of observation. An absorbed
blackbody was fitted to each of the spectra using \texttt{3ML}.
The priors of the free
parameters were chosen based on the \xmm\ fit results.
For the absorbing column
a Gaussian centered at $\mu = 2.7\times10^{21}$ cm$^{-2}$ and with a width of
$\sigma = 0.4\times 10^{21}$ cm$^{-2}$ was used.
The prior on $kT$ was a Gaussian with $\mu = 45$ eV and $\sigma = 4$ eV,
truncated at zero, and the prior on the normalization was a log-normal
distribution with $\mu$ = log(400) and $\sigma = 1$.
For the eROSITA data, the background was not modeled due to the low number
of counts; rather the data was binned to have at least 1 background
photon in every bin and a profile Poisson likelihood was used.
For the five epochs we obtain best-fit temperatures of
$kT_1 = 42^{+3}_{-2}$ eV,
$kT_2 = 44^{+3}_{-2}$ eV,
$kT_3 = 42^{+3}_{-2}$ eV,
$kT_4 = 42\pm2$ eV, and
$kT_5 = 43\pm2$ eV.
The corresponding
fluxes are listed in ED Tab. \ref{tabobsero}, and shown in ED
Fig. \ref{xraylc} together with the fluxes (or limits) of the other
X-ray missions.

\noindent\ul{ROSAT:}
\HP\ was originally identified\citep{HaberlPietsch1999} in a 8.3 ks
ROSAT/PSPC pointed observation
(ID: 500053p) of April 1992. We have re-analyzed
this observation, and find the source with a vignetting-corrected
count-rate of 0.005$\pm$0.001 PSPC cts/s (40$\pm$8 source counts).
A blackbody fit with free parameters leads to kT = 38$\pm$15 eV,
$N_{\rm H} = (0.9^{+3.2}_{-0.3}) \times 10^{21}$ cm$^{-2}$ and an unabsorbed
bolometric luminosity of $1.3^{+41.7}_{-1.0} \times 10^{36}$ erg/s.
A fit with a fixed, XMM-derived temperature
of 45 eV is statistically indistinguishable (due to the very small number of
counts and the low energy resolution), and results in an absorption-corrected
bolometric luminosity of $1.7^{+41}_{-1.0} \times 10^{36}$ erg/s,
consistent within the errors of the free fit.
A fit with fixed, XMM-derived temperature and $N_{\rm H}$ is substantially
worse.

\HP\ was not detected during the ROSAT all-sky survey, with
a PSPC count rate upper limit of $<$0.012 cts/s.
Using the best-fit spectral model of the above ROSAT pointed
observation leads to a luminosity limit of $<2.5\times 10^{36}$ erg/s,
while using the XMM-derived spectral parameters leads to
$<3.2\times 10^{37}$ erg/s. For consistency with the 
{\it Einstein} and EXOSAT upper limits we choose to plot the latter value
in Fig. \ref{xraylc}.

\noindent{\bf Arguments against an AM CVn interpretation:}

The He-dominated accretion disk and the NII and SiII lines
(ED Fig. \ref{fig:NIISII})
allow the possibility of an AM CVn nature of \HP.
However, a number of reasons argue against this interpretation:
(i) AM CVn objects have luminosities\citep{Ramsay+2018}
in the range of $10^{30}...10^{32}$ erg/s.
For this to be applicable to \HP, it would need to be at a distance of
order 100 pc.
(ii) This is incompatible with the Gaia data,
which suggest a minimum distance of 8--12 kpc.
(iii) Similarly, all AM CVn stars have large proper motion\citep{Ramsay+2018},
of order 0\farcs5/yr, due to their vicinity. This is a factor 100 larger
than that of \HP.
(iv) Finally, and most convincing, the velocity shift of all the strong
lines clearly indicates LMC membership. At that distance, an AM CVn
system is incompatible with the parameters we observe.

\noindent{\bf Comparison to known similar systems:}

To our knowledge, the only other 'known' system of this kind was the progenitor
of the He nova V445 Pup \cite{AshokBanerjee2003}.
A pre-outburst luminosity of log (L/L$_{\odot}$) = 4.34$\pm$0.36  would be
compatible with a 1.2--1.3 \msun\ star burning helium in a
shell\citep{Woudt+2009}. No optical spectrum
exists of the progenitor; the post-outburst spectra are H-deficient,
with the strongest lines being CII and FeII\citep{IijimaNakanishi2008}.
Based on photographic plates taken before the outburst,
an optical modulation
by a factor of 1.25 
and a period of 0.650654(10) days was found, and interpreted as
orbital variation of a common-envelope binary \citep{Goranskij+2010}.
There are 
three possibilities for the X-ray non-detection:
(i) the flux oscillations during burning with phases of low luminosity
\citep{Brooks+2016}, or
(ii) the substantial Galactic foreground absorption
in the case that the X-ray spectrum was similarly soft as \HP, or
(iii) an only slightly lower temperature as compared to \HP\ which would shift
the emission below the X-ray detection window.
Thus, the progenitor of the He nova V445 Pup could have been an
object similar to \HP.

\medskip

\noindent{\bf Data availability:}
Data from the ROSAT, \xmm, Swift and TESS missions as well as from the
OGLE and MACHO projects are publicly available.
eROSITA data of the first survey (eRASS1) of HP99-159 will be made public
as part of the eRASS1 data release, presently scheduled for March 2023.
Data of the subsequent eROSITA surveys (eRASS2 and later) will be made public
according to the plan as provided at https://erosita.mpe.mpg.de/erass/.
The optical spectra taken with the SALT telescope are available
at https://cloudcape.saao.ac.za/index.php/s/g8M1q1ya8ef7Fzd.

\noindent{\bf Code availability:}
All data analysis code is publicly available, as referenced in the text.

\newpage
\renewcommand{\refname}{Additional References}
{}

\newpage

\noindent{\bf Acknowledgements:}

RW is supported by the German Science Foundation (DFG) under contract
GR1350/17-1, and JMB by the DFG-funded Collaborative Research Center SFB 1258.
We thank M. Freyberg (MPE Garching) for discussion of the \xmm\
background modelling, and S. Rappaport (MIT) on TESS time-series analysis.
JG is grateful to the Swift team for the rapid scheduling of the UVOT
observation.

This work is based on data from eROSITA, the soft X-ray instrument aboard SRG, a joint Russian-German science mission supported by the Russian Space Agency (Roskosmos), in the interests of the Russian Academy of Sciences represented by its Space Research Institute (IKI), and the Deutsches Zentrum f\"ur Luft- und Raumfahrt (DLR). The SRG spacecraft was built by Lavochkin Association (NPOL) and its subcontractors, and is operated by NPOL with support from the Max-Planck Institute for Extraterrestrial Physics (MPE).
The development and construction of the eROSITA X-ray instrument was led by MPE, with contributions from the Dr. Karl Remeis Observatory Bamberg \& ECAP (FAU Erlangen-N\"urnberg), the University of Hamburg Observatory, the Leibniz Institute for Astrophysics Potsdam (AIP), and the Institute for Astronomy and Astrophysics of the University of T\"ubingen, with the support of DLR and the Max-Planck Society. The Argelander Institute for Astronomy of the University of Bonn and the Ludwig Maximilians Universit\"at Munich also participated in the science preparation for eROSITA.
The eROSITA data shown here were processed using the eSASS software system developed by the German eROSITA consortium.

This paper utilizes public domain data obtained by the MACHO Project, jointly funded by the US Department of Energy through the University of California, Lawrence Livermore National Laboratory under contract No. W-7405-Eng-48, by the National Science Foundation through the Center for Particle Astrophysics of the University of California under cooperative agreement AST-8809616, and by the Mount Stromlo and Siding Spring Observatory, part of the Australian National University.

Some of the observations reported in this paper were obtained with the Southern African Large Telescope (SALT) under program 2018-2-LSP-001.

This paper includes data collected by the TESS mission. Funding for the TESS mission is provided by NASA's Science Mission Directorate. Resources used in this work were provided by the NASA High-End Computing (HEC) Program through the NASA Advanced Supercomputing (NAS) Division at Ames Research Center for the production of the SPOC data products.

\noindent{\bf Author contributions}

CM, FH, RW and PM analyzed the X-ray data, with JMB providing the
3ML environment. DAHB and IMM obtained the SALT spectra,  AU
provided the OGLE data, and RJ and RV analyzed the TESS data.
JG recognized the He burning nature, and with NL, HR, JB and KW
derived the binary system constraints.
All authors contributed to the scientific discussion and the writing
of the manuscript.

\noindent{\bf Additional Information}

The authors declare no competing interests.

Correspondence and requests for materials should be addressed to JG
(jcg@mpe.mpg.de).

Reprints and permissions information is available at www.nature.com/reprints.

\newpage

\captionsetup[table]{name=Extended Data Table, labelsep=bar}
\captionsetup[figure]{name=Extended Data Figure, labelsep=bar}
\setcounter{figure}{0}
\setcounter{table}{0}

\section*{Extended Data}


\bigskip\bigskip

\begin{table}[ht]
\centering
\caption[]{{\bf Peaks in the Lomb-Scargle periodogram of the OGLE data.}}
\begin{tabular}{lcccccc}
\hline\hline\noalign{\smallskip}
Peak & Maximum power & P & f & 1.0 + f & 1.0 - f & f - f$_{\rm III}$ \\
No. &        & (days) & (1/days) & (1/days) & (1/days) & (1/days) \\
\noalign{\smallskip}\hline\noalign{\smallskip}
I   & 0.6367 &  1.1635 & 0.8595 & 1.8595 & 0.1405 & 0.7190 \\
II  & 0.4733 &  6.9794 & 0.1433 & 1.1433 & 0.8567 & 2.772$\times 10^{-3}$\\
III & 0.3271 &  7.1171 & 0.1405 & 1.1405 & 0.8595 & -- \\
\noalign{\smallskip}\hline\smallskip
\end{tabular}

\noindent 
{\bf Notes:} (i) 1/(f$_{\rm II}$ - f$_{\rm III}$) = 360.8 days indicates that these two frequencies are one year aliases of each other.
    (ii)  1 -- f$_{\rm III}$ = 0.859494 $\approx$ f$_{\rm I}$, suggesting that 
     f$_{\rm I}$ and f$_{\rm III}$ are one day aliases.
     (iii) Possible shorter periods are: 1/(1+f$_{\rm III}$) = 0.87680 days, or
     1/(1+f$_{\rm I}$) = 0.53778 days.

\label{tab:OGLE_f}
\end{table}

\newpage

\begin{table}[h]
\caption{{\bf X-ray Observations of \HP . }} 
\begin{tabular}{ccccc} 
\hline\hline\noalign{\smallskip}
  Mission & observation time & exposure & count rate$^{a}$  & luminosity$^{a}$ \\
          & T$_{start}$ -- T$_{stop}$ (UTC) & (ks) & (cts s$^{-1}$) & (erg s$^{-1}$)\\
  \noalign{\smallskip}\hline\noalign{\smallskip}
{\it Einstein}& 1979-04-12 17:10--1979-04-12 17:50 & 2.4 & $<$0.01 & $<$1.8$\times 10^{38}$\\
EXOSAT       &  1986-01-21 04:17--1986-01-21 12:00 & 27.8~~& $<0.0047$ & $<$9.8$\times 10^{37}$  \\
ROSAT Surv.  &  1990-07-12 01:33--1990-07-16 02:54 & 2.4 & $<$0.012 & $<$3.2$\times 10^{37}$\\
ROSAT 500053p&  1992-04-09 16:10--1992-04-13 19:15 & 8.3 & 0.005$\pm0.001$ & $1.7^{+41}_{-1.0} \times 10^{36}$ \\
XMM-Newton   &  2019-09-16 18:35--2019-09-17 03:25 & 28.6~~ & 0.056$\pm$0.003 & $6.8^{+7.0}_{-3.5} \times 10^{36}$ \\
eRASS0/1$^{b}$   & 2019-12-11 23:33--2019-12-27 15:33 & 1.2$^{c}$ & 0.10$\pm0.01$$^{d}$& $7.1^{+4.5}_{-2.7} \times 10^{36}$\\
eRASS1/2$^{b}$   & 2020-06-07 00:57--2020-06-23 20:57 & 1.0$^{c}$ & 0.10$\pm0.01$$^{d}$ & $6.5^{+4.2}_{-2.6} \times 10^{36}$\\
eRASS2/3$^{b}$   & 2020-12-13 14:33--2020-12-28 14:33 & 0.9$^{c}$ & 0.06$\pm0.01$$^{d}$& $7.1^{+4.5}_{-2.7} \times 10^{36}$\\ 
eRASS3/4$^{b}$   & 2021-06-11 13:57--2021-06-27 05:57 & 1.1$^{c}$ & 0.09$\pm0.01$$^{d}$& $6.5^{+3.9}_{-2.6} \times 10^{36}$\\
eRASS4/5$^{b}$   & 2021-12-17 15:33--2021-12-19 15:33 & 0.1$^{c}$ & 0.17$\pm0.08$$^{d}$& $8.0^{+4.8}_{-3.0} \times 10^{36}$\\
\hline
\label{tabobsero} 
\end{tabular} 

\noindent
(a) Count rate is in the 0.2--2\,keV band, and luminosity is
   foreground-absorption corrected.\\
(b) eROSITA test scans from 2019-12-08 to 2019-12-11 are designated as eRASS0.
   The source position was covered by these test scans and in the early phase
   of eRASS1. The $\approx$2 week visibility of \HP\ for eROSITA
   starts at the end of each formal eRASS survey and extends into the start
   of the following eRASS: we group this into 'epochs' of continuous coverage.
   \\
(c) Net exposure per telescope after correcting for vignetting, averaged
   over the 5 telescope modules used. \\
(d) Net source count rate after correcting for vignetting, summed over
   the 5 telescope modules used.
\end{table}

\newpage

\begin{figure}[ht]
  \centering
\includegraphics[width=0.6\textwidth]{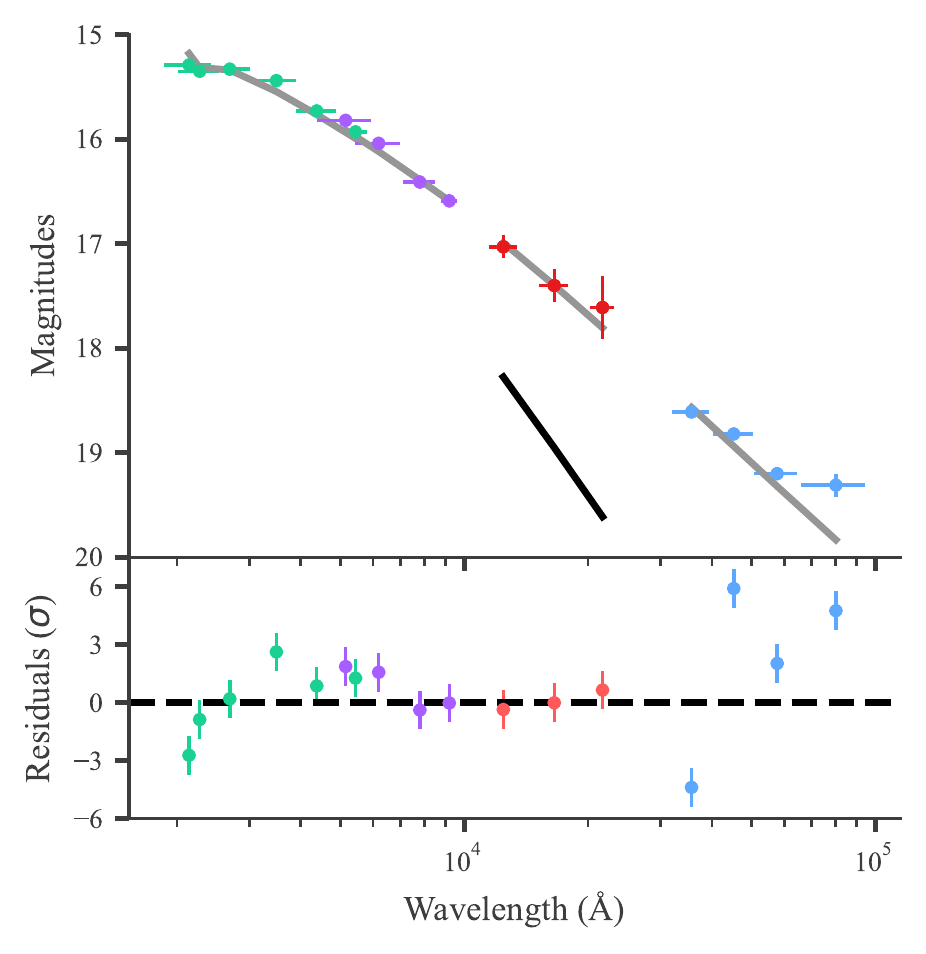}
\caption{{\bf Spectral energy distribution of \HP.}
  The collection of non-simultaneous photometry of \HP\
  from SkyMapper (griz; purple), 2MASS (JHK; red),
  and Spitzer (IRAC-1 to IRAC-4; blue), together with near-simultaneous
  5-filter UV/optical photometry with Swift/UVOT (green).
  Error bars are at the 1$\sigma$ level.
  The extinction has been fit by forward-folding a powerlaw
  and allowing for the sum of Milky Way and LMC dust extinction curves
  (best-fit E(B-V) = 0.14$\pm$0.01).
  The orbital variability amplitude of $\pm$0.15 mag is the likely reason
  for the scatter in the non-simultaneous data, but has not been included
  as ``systematic'' error in the fitting.
  The slope of the observed spectral energy distribution
  (grey line; $F_\nu \propto \nu^{1.48\pm0.02}$)
  is clearly different from that expected from a non-irradiated
  accretion disk (black line: $F_\nu \propto \nu^{1/3}$),
  consistent with the properties
  of canonical supersoft X-ray sources\citep{PophamDiStefano1996}.
}
\label{fig:SED}
\end{figure}


\begin{figure}[ht]
  \centering
  \includegraphics[width=0.85\textwidth]{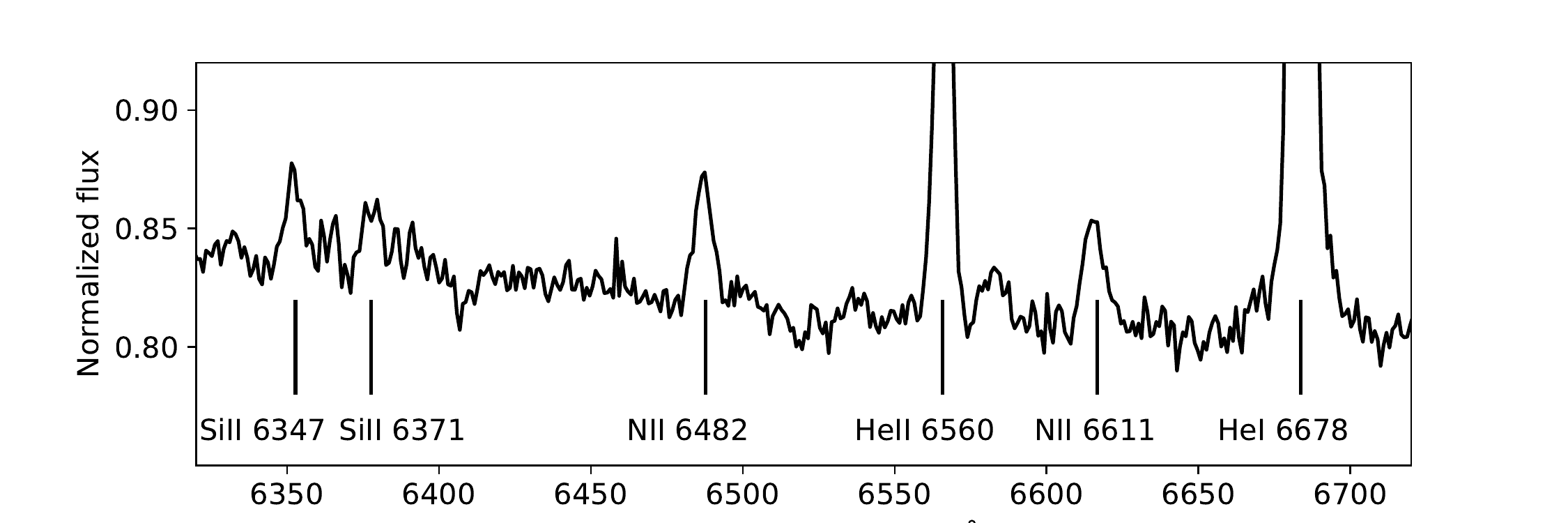}
  \caption{{\bf Faint Nitrogen and Silicon lines.}
    Zoom-in of the low-resolution SALT spectrum, showing four of the
    seven NII and SiII emission lines identified in addition to the
    He lines.
  }
  \label{fig:NIISII}
\end{figure}

\begin{figure}[ht]
  \includegraphics[width=0.95\textwidth]{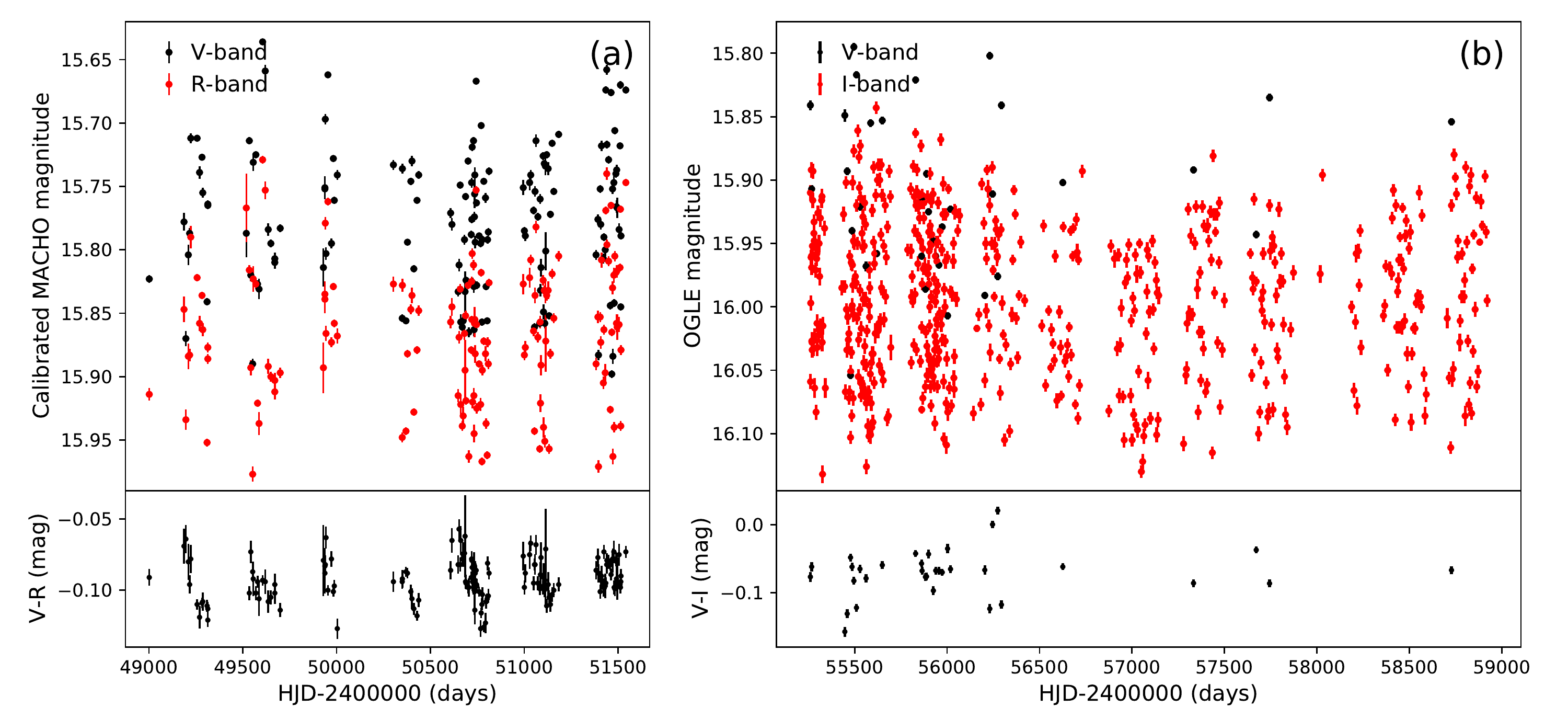}
  \caption{{\bf Optical long-term light-curve of \HP.}
    MACHO (panel (a); Jan. 1993 -- Jan. 2000) and
    OGLE I-band ((b); Mar. 2010 -- Mar. 2020) light curve of \HP.
    Error bars are at the 1$\sigma$ level.
    The MACHO color has an additional systematic error of $\pm$0.028 mag.
    The $V$-magnitude offset between the two panels is
    due to the accuracy of the absolute photometric calibration of the
    MACHO data.
  }
  \label{fig:machooglelc}
\end{figure}

\newpage

\begin{figure}[ht]
   \centering
   \includegraphics[width=0.7\textwidth]{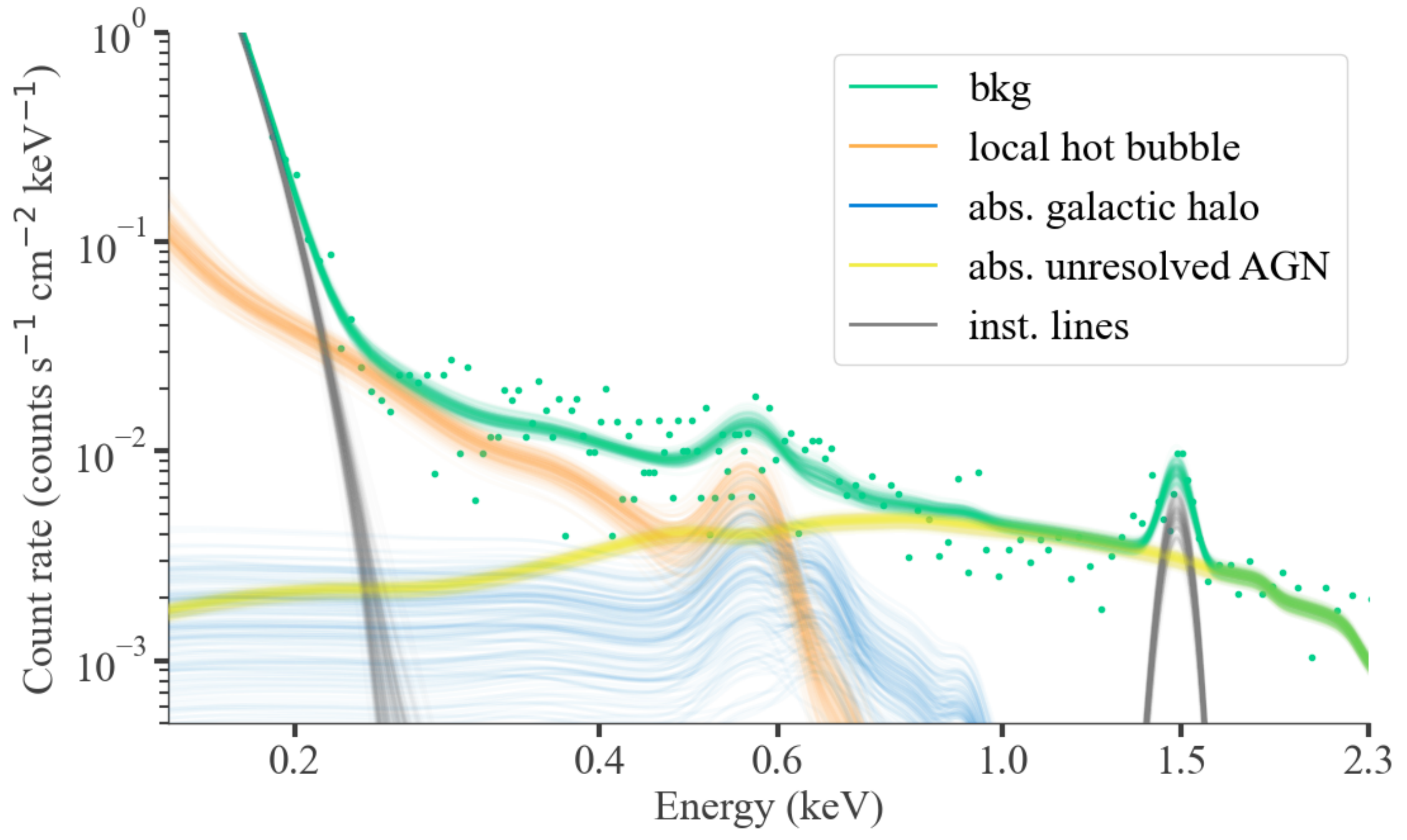}
   \caption{{\bf \xmm\ X-ray background modelling.}
     The \xmm\ X-ray background is extracted from a distinct region
     on the detector, but located close to the source region, and
     in a first step separately modelled as the sum of four components,
     as labelled. During the spectral fit of the photons from the
     source extraction region, the background spectrum is scaled by
     the ratio of the extraction areas, and the parameters of three of
     the four background model components were linked; for the read-noise
     component (steeply falling grey component below 0.3 keV), due to
     its strong variation over the detector, the parameters were left
     free.
   }
  \label{fig:Xraybkg}
\end{figure}

\newpage

\begin{figure}[th]
  \centering
  \includegraphics[width=14.cm]{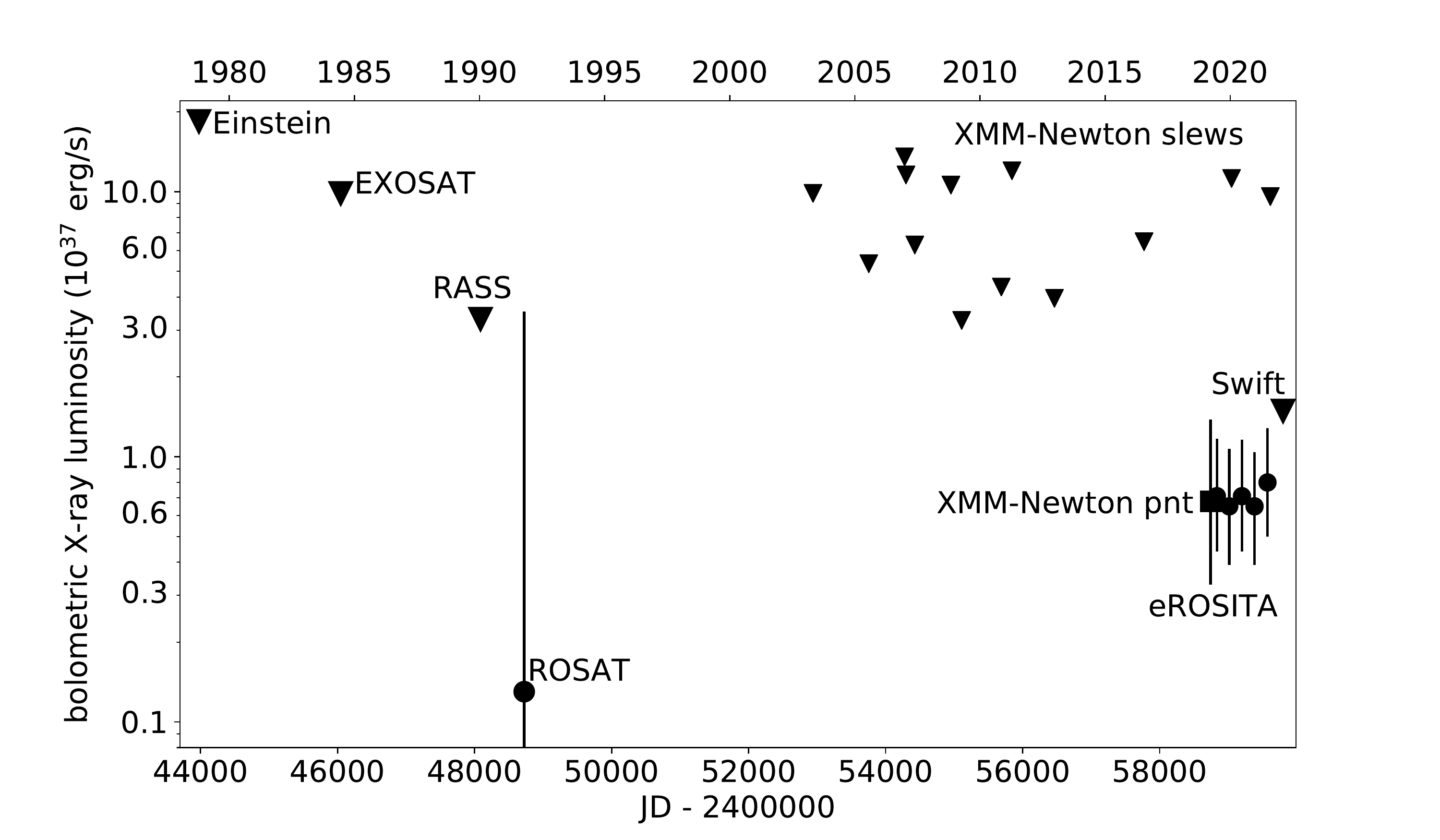}
   \vspace{-0.3cm}
  \caption[X-ray lc]{{\bf Long-term X-ray light curve.}
    Compilation of X-ray 2$\sigma$ upper limits (downward triangles)
    and detections
    (filled circles with 1$\sigma$ error bars) from previous X-ray missions. 
    The upper limits from {\it Einstein}, EXOSAT and the \xmm\ slews
    have been computed
    using \texttt{http://xmmuls.esac.esa.int/upperlimitserver/}
    with kT=60 eV and $N_{\rm H}$ = 10$^{21}$ cm$^{-2}$, and then transformed
    to the best-fit \xmm\ spectral values using WebPIMMS
    (\texttt{https://heasarc.gsfc.nasa.gov/cgi-bin/Tools/w3pimms/w3pimms.pl}).
  \label{xraylc}}
\end{figure}


\newpage


\begin{thebibliography}{}
  \renewcommand{\baselinestretch}{0.005}
  \onehalfspacing
\itemsep-.2ex

\bibitem[1]{Elias+1985} Elias, J.H., Matthews, K., Neugebauer, G.,
  Persson, S.E., Type I supernovae in the infrared and their use as
  distance indicators.
   {\it Astrophys. J.} {\bf 296}, 379--389  (1985)

\bibitem[2]{Riess+1996} Riess, A.G., Press, W.H., Kirshner, R.P.,
  A Precise Distance Indicator: Type IA Supernova Multicolor Light-Curve Shapes.
   {\it Astrophys. J.} {\bf 473}, 88--109 (1996) 

\bibitem[3]{Hoyle1946} Hoyle, F.,
  The synthesis of the elements from hydrogen.
  {\it Mon. Not. R. Astron. Soc.} {\bf 016}, 343--383 (1946)

\bibitem[4]{Burbidge+1957}  Burbidge, E.M., Burbidge, G.R., Fowler, W.A.,
  Hoyle, F., Synthesis of the Elements in Stars.
  {\it Rev. Mod. Phys.} {\bf 29}, 547–-650 (1957)

\bibitem[5]{Greiner2000} Greiner, J., Catalog of supersoft X-ray sources.
  {\it New Astron.} {\bf 5}, 137--141 (2000)

\bibitem[6]{WhelanIben1973} Whelan, J., Iben, I. Jr,
      Binaries and Supernovae of Type I.
  {\it Astrophys. J.} {\bf 186}, 1007 (1973)
  
\bibitem[7]{IbenTutukov1994} Iben, I. Jr., Tutukov, A.V.,
  Helium-accreting Degenerate Dwarfs as Presupernovae and Scenarios for the Ultrasoft X-Ray Sources.
  {\it Astrophys. J.} {\bf 431}, 264--272 (1994)

\bibitem[8]{YoonLanger2003} Yoon, S.-C., Langer, N.,
  The first binary star evolution model producing a Chandrasekhar mass white dwarf.
   {\it Astron. Astrophys.} {\bf 412}, L53--L56 (2003)

\bibitem[9]{Wang+2009} Wang, B., Meng, X., Chen, X., Han, Z.
  The helium star donor channel for the progenitors of Type Ia supernovae.
  {\it Mon. Not. R. Astron. Soc.} {\bf 395}, 847--854 (2009)

\bibitem[10]{Wheeler+1975} Wheeler, J.C., Lecar, M., McKee, C.F.,
Supernovae in binary systems.
   {\it Astrophys. J.} {\bf 200}, 145--157 (1975)
  
\bibitem[11]{Kawai+1988} Kawai, Y., Saio, H., Nomoto, K.,
  Steady State Models of White Dwarfs Accreting Helium or Carbon/Oxygen--rich Matter.
  {\it Astrophys. J.} {\bf 328}, 207--212 (1988) 
  
\bibitem[12]{IbenTutukov1989} Iben, I. Jr., Tutukov, A.V.,
  Model Stars with Degenerate Dwarf Cores and Helium-burning Shells: A Stationary-burning Approximation.
  {\it Astrophys. J.} {\bf 342}, 430--448 (1989)

\bibitem[13]{Foley+2013}  Foley, R.J., Challis, P.J., Chornock, R.,
  Ganeshalingam, M., Li, W., et al.
  Type Iax Supernovae: A New Class of Stellar Explosion.
  {\it Astrophys. J.} {\bf 767}, id. 57 (2013)
  
\bibitem[14]{HaberlPietsch1999} Haberl F., Pietsch W.,
  A ROSAT PSPC catalogue of X-ray sources in the LMC region.
  {\it Astron. Astrophys. Suppl. Ser.} {\bf 139}, 277 (1999)

\bibitem[15]{vanderMarel+2002} van der Marel, R.P., Alves, D.R., Hardy, E.,
  Suntzeff, N.B. 
  New Understanding of Large Magellanic Cloud Structure, Dynamics, and Orbit from Carbon Star Kinematics.
  {\it Astron. J.} {\bf 124}, 2639--2663 (2002)

\bibitem[16]{Pietrzynski+2019} Pietrzynski, G., Graczyk, D., Gallenne, A.
  A distance to the Large Magellanic Cloud that is precise to one per cent.
  {\it Nature} {\bf 567}, 200--203 (2019)
  
\bibitem[17]{Greiner+1991}  Greiner, J., Hasinger, G., Kahabka, P.
  ROSAT observation of two supersoft sources in the Large Magellanic Cloud.
  {\it Astron. Astrophys.} {\bf 246},  L17--L20 (1991)
  
\bibitem[18]{Heuvel+1992} van den Heuvel, E.P.J., Bhattacharya, D.,
  Nomoto, K., Rappaport, S.A.,
  Accreting white dwarf models for CAL 83, CAL 87 and other ultrasoft X-ray sources in the LMC.
  {\it Astron. Astrophys.} {\bf 262}, 97--105 (1992)

\bibitem[19]{Nomoto1982} Nomoto, K.
  Accreting white dwarf models for type I supernovae. I. Presupernova evolution and triggering mechanisms.
  {\it Astrophys. J.} {\bf 253}, 798--810 (1982)

\bibitem[20]{Fujimoto1982} Fujimoto, M.Y.
  A Theory of Hydrogen Shell Flashes on Accreting White Dwarfs. II. The Stable Shell Burning and the Recurrence Period of Shell Flashes.
  {\it Astrophys. J.} {\bf 257}, 767--779 (1982)

\bibitem[21]{Yoon+2004} Yoon, S.-C., Langer, N., Scheithauer, S.,
  Effects of rotation on the helium burning shell source in accreting white dwarfs.
   {\it Astron. Astrophys.} {\bf 425}, 217--228 (2004) 

\bibitem[22]{Piersanti+2014} Piersanti, L., Tornamb\'e, A., Yungelson L.R.,
  He-accreting white dwarfs: accretion regimes and final outcomes.
  {\it Mon. Not. R. Astron. Soc.} {\bf 445}, 3239–-3262 (2014)
  
\bibitem[23]{Wong+2021} Wong T.L.S., Schwab, J., G\"otberg, Y.
  Pre-Explosion Properties of Helium Star Donors to Thermonuclear Supernovae.
   {\it Astrophys. J.} {\bf 922}, id. 241 (2021)
   
\bibitem[24]{YoonLanger2004} Yoon, S.-C., Langer, N.,
  Helium accreting CO white dwarfs with rotation: Helium novae instead of double detonation.
  {\it Astron. Astrophys.} {\bf 419}, 645--652 (2004)
   
\bibitem[25]{Brooks+2016} Brooks, J., Bildsten, L., Schwab, J., Paxton, B.,
   Carbon Shell or Core Ignitions in White Dwarfs Accreting from Helium Stars.
   {\it Astrophys. J.} {\bf 821}, id. 28 (2016)

\bibitem[26]{WongSchwab2019} Wong, T.L.S., Schwab, J.,
  Evolution of Helium Star-White Dwarf Binaries Leading up to Thermonuclear Supernovae.
  {\it Astrophys. J.} {\bf 878}, id. 100 (2019)

\bibitem[27]{KatoHachisu1999} Kato, M., Hachisu, I.
A New Estimation of Mass Accumulation Efficiency in Helium Shell Flashes toward Type IA Supernova Explosions.
  {\it Astrophys. J.} {\bf 513}, L41 (1999)

\bibitem[28]{Langer+1989} Langer, N.
  Standard models of Wolf-Rayet stars.
   {\it Astron. Astrophys.} {\bf 210}, 93--113 (2019)

\bibitem[29]{DelgadoThomas1981} Delgado, A.J., Thomas, H.-C.
  Mass transfer in a binary system - The evolution of the mass-giving helium star.
  {\it Astron. Astrophys.} {\bf 96}, 142--145 (1981)
   
\bibitem[30]{Liu+2013} Liu, Z.-W., Pakmor, R., Seitenzahl, I.R.,
  Hillebrandt, W., Kromer, M. et al.
The Impact of Type Ia Supernova Explosions on Helium Companions in the Chandrasekhar-mass Explosion Scenario.
 {\it Astrophys. J.} {\bf 774}, id. 37 (2013)

\bibitem[31]{Kromer+2013}  Kromer, M., Fink, M., Stanishev, V.,
  Taubenberger, S., Ciaraldi-Schoolman, F., et al.
  3D deflagration simulations leaving bound remnants: a model for 2002cx-like Type Ia supernovae.
  {\it Mon. Not. R. Astron. Soc.} {\bf 429}, 2287--2297 (2013)
 
\bibitem[32]{Zeng+2020} Zeng, Y., Liu, Z.-W., Han, Z.
  The Interaction of Type Iax Supernova Ejecta with a Helium Companion Star.
  {\it Astrophys. J.} {\bf 898}, id. 12 (2020)

\bibitem[33]{McCully+2014}  McCully, C., Jha, S.W., Foley, R.J., Bildsten, L.,
  Fong, W.-F., et al.
  A luminous, blue progenitor system for the type Iax supernova 2012Z.
  {\it Nature} {\bf 512}, 54--56 (2014)

\bibitem[34]{Kool+2022} Kool, E.C., Johansson, J., Sollerman, J., Mold\'on, J.,
   Moriya, T.J., et al., 
   A radio-detected thermonuclear supernova from a single-degenerate progenitor with a helium star donor.
  {\it Nature} (subm.; arXiv:2210.07725) (2022)

\bibitem[35]{Rotondo2011} Rotondo, M., Rueda, J.A., Ruffini, R., Xue, S.-S.
  Relativistic Feynman-Metropolis-Teller theory for white dwarfs in general relativity.
  {\it Phys. Rev. D} {\bf 84}, id. 084007 (2011)


\end{thebibliography}

\begin{thebibliography}{}
  \renewcommand{\baselinestretch}{0.005}
  \onehalfspacing
\itemsep-.2ex

\bibitem[36]{Onken+2019} Onken, C.A., Wolf, C., Bessel, M.S. et al.
  SkyMapper Southern Survey: Second data release (DR2).
  {\it Publ. Astron. Soc. Aust.} {\bf 36}, id. e033 (2019)

\bibitem[37]{Smak1989} Smak, J., On the M$_V - \dot M$ relation for accretion
  disks in cataclysmic binaries. {\it Acta Astron.} {\bf 30}, 317--321 (1989)
  
\bibitem[38]{Skowron+21} Skowron, D.M., Skowron, J., Udalski A. et al.
  OGLE-ing the Magellanic System: Optical Reddening Maps of the Large and Small Magellanic Clouds from Red Clump Stars.
  {\it Astrophys. J. Suppl. Ser.} {\bf 252}, 23 (2021)

\bibitem[39]{Crampton+1987}  Crampton, D., Cowley, A.P., Hutchings, J.B.,
  Schmidtke, P.C., Thompson, I.B., Liebert, J.
  CAL 83: A Puzzling X-Ray Source in the Large Magellanic Cloud.
  {\it Astrophys. J.} {\bf 321}, 745--754 (1987)

\bibitem[40]{PophamDiStefano1996} Popham, R., DiStefano, R.
  Accretion disks in supersoft X-ray sources.
  {\it Lect. Notes in Phys.} {\bf 472}, 65--72 (1996)
  
\bibitem[41]{Udalski+2008} Udalski, A., Szymanski, M.K., Soszynski, I.,
  Poleski, R.
  The Optical Gravitational Lensing Experiment. Final Reductions of the OGLE-III Data.
  {\it Acta Astron.} {\bf 58}, 69--87 (2008)

\bibitem[42]{Udalski+2015} Udalski, A., Szymanski, M.K., Szymanski, G.
OGLE-IV: fourth phase of the optical gravitational lensing experiment.
  {\it Acta Astron.} {\bf 65}, 1--38 (2015)

\bibitem[43]{Alcock+1999} Alcock, C., Allsman, R.A., Alves, D.R.,
  Axelrod, T.S., Becker, A.C., et al.
Calibration of the MACHO Photometry Database.
  {\it Publ. Astron. Soc. Pac.} {\bf 111}, 1539--1558 (1999)

\bibitem[44]{Greiner+2002} Greiner, J., Di\,Stefano, R.
  X-ray off states and optical variability in CAL 83.
  {\it Astron. Astrophys.} {\bf 387}, 944--954 (2002)

\bibitem[45]{Kim+2014} Kim, D.-W., Protopapas, P., Bailer-Jones, C.A.L., et al.
  The EPOCH Project. I. Periodic variable stars in the EROS-2 LMC database.
  {\it Astron. Astrophys.} {\bf 566}, A43 (2014)

\bibitem[46]{Ricker+2014} Ricker, G.R., Winn, J.N., Vanderspek, R. et al.
  Transiting Exoplanet Survey Satellite (TESS).
  {\it J. Astron. Tel., Instr \& Sys.} {\bf 1}, id. 014003

\bibitem[47]{Young+1972} Young, A., Nelson, B., Mielbrecht, R.
  An Old Evolved Binary in the Galactic Halo.
  {\it Astrophys. J.} {\bf 174}, 27--31 (1972)
  
\bibitem[48]{Feinstein+2019} Feinstein, A.D., Montet, B.T.,
  Foreman-Mackey, D. et al.
  eleanor: An Open-source Tool for Extracting Light Curves from the TESS Full-frame Images.
  {\it Publ. Astron. Soc. Pac.} {\bf 131}, 094502
 
\bibitem[49]{Burgh+2003} Burgh, E.B., Nordsieck, K.H., Kobulnicky, H.A.,
  et al.
Prime Focus Imaging Spectrograph for the Southern African Large Telescope: optical design.
  {\it Proc. SPIE} {\bf 4841}, 1463--1471 (2003)

\bibitem[50]{Crause+2014} Crause, L.A., Sharples, R.M., Bramall, D.G., et al.
  Performance of the Southern African Large Telescope  High Resolution Spectrograph (HRS).
  {\it Proc. SPIE} {\bf 9147}, id. 91476T (2014)

\bibitem[51]{Crawford2015} Crawford, S.M.
  pyhrs: Spectroscopic data reduction package for SALT.
  {\it Astrophys. Source Code Lib.} ascl:1511.005 (2015)

\bibitem[52]{Vianello+2017}  Vianello, G., Lauer, R.J., Burgess, J.M.
  The Multi-Mission Maximum Likelihood framework (\texttt{3ML}).
  in {\it Proc. of the 7th Interntl. Fermi Symp.}, SISSA IFS2017, id. 130
  \texttt{https://pos.sissa.it/cgi-bin/reader/conf.cgi?confid=312}, (2017)

\bibitem[53]{Burgess+2021}  Burgess, J.M., Fleischhack, H., Vianello, G.,
  et al. The Multi-Mission Maximum Likelihood framework (\texttt{3ML}).
  {\it Zenodo} 5646954, doi:10.5281/zenodo.5646954 (2021)

\bibitem[54]{Hansen+2004} Hansen, C.J., Kawaler, S.D., Trimble, V.
  Stellar interiors: Physical principles, structure, and evolution
  2nd ed., New York: Springer, (2004)

\bibitem[55]{RauchWerner2010} Rauch, T., Werner, K.,
  Non-LTE model atmospheres for supersoft X-ray sources.
  {\it Astron. Nachr.} {\bf 331}, 146--151 (2010)
  
\bibitem[56]{Predehl+2021} Predehl, P., Andritschke R., Arefiev, V., et al.
  The eROSITA X-ray telescope on SRG.
  {\it Astron. Astrophys.} {\bf 647}, A1 (2021)

\bibitem[57]{Brunner+2022} Brunner, H., Liu, T., Lamer, G., et al.
The eROSITA Final Equatorial Depth Survey (eFEDS). X-ray catalogue.
  {\it Astron. Astrophys.} {\bf 661}, A1 (2022)
  
\bibitem[58]{Ramsay+2018} Ramsay G., Green, M.J., Marsh, T.R., et al.
  Physical properties of AM CVn stars: New insights from Gaia DR2.
  {\it Astron. Astrophys.} {\bf 620}, A141 (2018)
  
\bibitem[59]{AshokBanerjee2003} Ashok, N.M., Banerjee, D.P.K.,
   The enigmatic outburst of V445 Puppis - A possible helium nova? 
   {\it Astron. Astrophys.} {\bf 408}, 1007--1015 (2003) 
   
\bibitem[60]{Woudt+2009} Woudt, P.A., Steeghs, D., Karovska, M., et al.,
  The Expanding Bipolar Shell of the Helium Nova V445 Puppis.
  {\it Astrophys. J.} {\bf 709}, 738--746 (2009)

\bibitem[61]{IijimaNakanishi2008} Iijima, T., Nakanishi, H.
  Spectroscopic observations of the first helium nova V445 Puppis.
  {\it Astron. Astrophys.} {\bf 482}, 865--877 (2008)
  
\bibitem[62]{Goranskij+2010}  Goranskij, V., Shugarov, S., Zharova, A.,
  Kroll, P., Barsukova, E.A.
  The Progenitor and Remnant of the Helium Nova V445 Puppis.
  {\it Perem. Zvezdy} {\bf 30}, No. 4 (2009)


\end{thebibliography}
\end{document}